\documentclass{article}
\usepackage[doublespacing]{setspace}
\usepackage{gensymb}
\usepackage{authblk}
\usepackage{graphicx}
\usepackage[top=2cm,bottom=2cm,left=2cm,right=2cm]{geometry}
\graphicspath{{../figs/}}

\title{Intrinsic Strain Aging, $\Sigma 3$  Boundaries, and Origins of Cellular Substructure in Additively Manufactured 316L}

\author[1]{Andrew J. Birnbaum}
\author[1]{John C. Steuben}
\author[2]{Erin J. Barrick}
\author[1]{Athanasios P. Iliopoulos}
\author[1]{John G. Michopoulos}
\affil[1]{United States Naval Research Laboratory, Materials Science Division, Washington, DC, USA.}
\affil[2]{Lehigh University, Dept.of Material Science and Engineering, Bethlehem, PA, USA.}

\begin{document}

\maketitle

\begin{abstract}
The observation of sub-grained cellular features in additively manufactured (AM)/selectively laser melted (SLM) 316L stainless steel components has remained an interesting, though incompletely understood phenomenon. However, the recently observed correlation linking the presence of these features with significantly enhanced mechanical strength in SLM 316L materials \cite{Wang201863} has driven a renewed interest and effort toward elucidating the mechanism(s) by which they are formed. To date, the dominant hypothesis, cellular solidification followed by dislocation-solute entanglement \cite{Saeidi2015221}, remains incompatible with the ensemble of reported observations from multiple independent studies. This effort offers direct evidence of a previously unrecognized interaction of phenomena, that, when acting in concert, give rise to this commonly observed substructure. These phenomena include SLM-induced intrinsic strain-aging, Cottrell atmosphere formation, and twin-boundary enhanced mass diffusion to structural defects.
\end{abstract}

\section*{Introduction}
While all research groups studying SLM of 316L report observing cellular substructures via some combination of optical, scanning and/or transmission electron microscopy (SEM and TEM, respectively), only a subset of these groups report the presence of an increased concentration of certain elements (typically Cr and/or Mo) to  cell walls via TEM/EDS  \cite{Wang201863, Liu2018354, Saeidi2015463} (Energy Dispersive X-Ray Spectroscopy) or atom probe tomography (APT) \cite{Krakhmalev2018}. However, despite similar methodologies, a different subset of research groups do not observe the presence of such chemical heterogeneity  \cite{Qiu2018, Saeidi2015221}. Interestingly, two parallel studies performed by the above-mentioned group \cite{Wang201863} resulted in observation of microsegregation in one study, but not the other. There is currently no consensus explanation for this acknowledged inconsistency \cite{Prashanth201727, Tucho2018910, saeidi2013606}. 

In contrast to the intermittent observation of solute concentration, TEM inspection shows that cell walls are \textit{always} associated with the presence of dense dislocation tangles. These dislocation structures represent one echelon in the above-referenced microstructural “hierarchy” identified as the basis for enhanced strength and ductility \cite{Wang201863, Liu2018354}. The existing theory for cell formation invokes cellular solidification as the primary mechanism for the formation of these arrangements. It has thus been posited that the \textit{compositional} cell structure acts as a spatial framework for pinning mobile dislocations as they propagate due to the generation of residual stress upon cooling  \cite{Wang201863}.

All studies that observe the presence or absence of microsegregation characterize samples obtained from macro-scale AM structures, i.e. components that consist of multiple laser scans in-plane, as well as multiple layers of processed materials. In order to eliminate the confounding variables associated with multiple heat/re-heat and melt/re-melt cycles, we have opted to focus on single SLM tracks processed on a 316L base plate with a single layer of powder. Several single-track efforts have been conducted previously, specifically on 316L \cite{saeidi2013606, Yadroitsev2012201, Krakhmalev201712, Liu2017}, though none characterize spatial compositional variation at the requisite scale/resolution.

\section*{The Presence and Absence of Surface Substructure}

\begin{figure}[h]
	\includegraphics[width=0.99\linewidth]{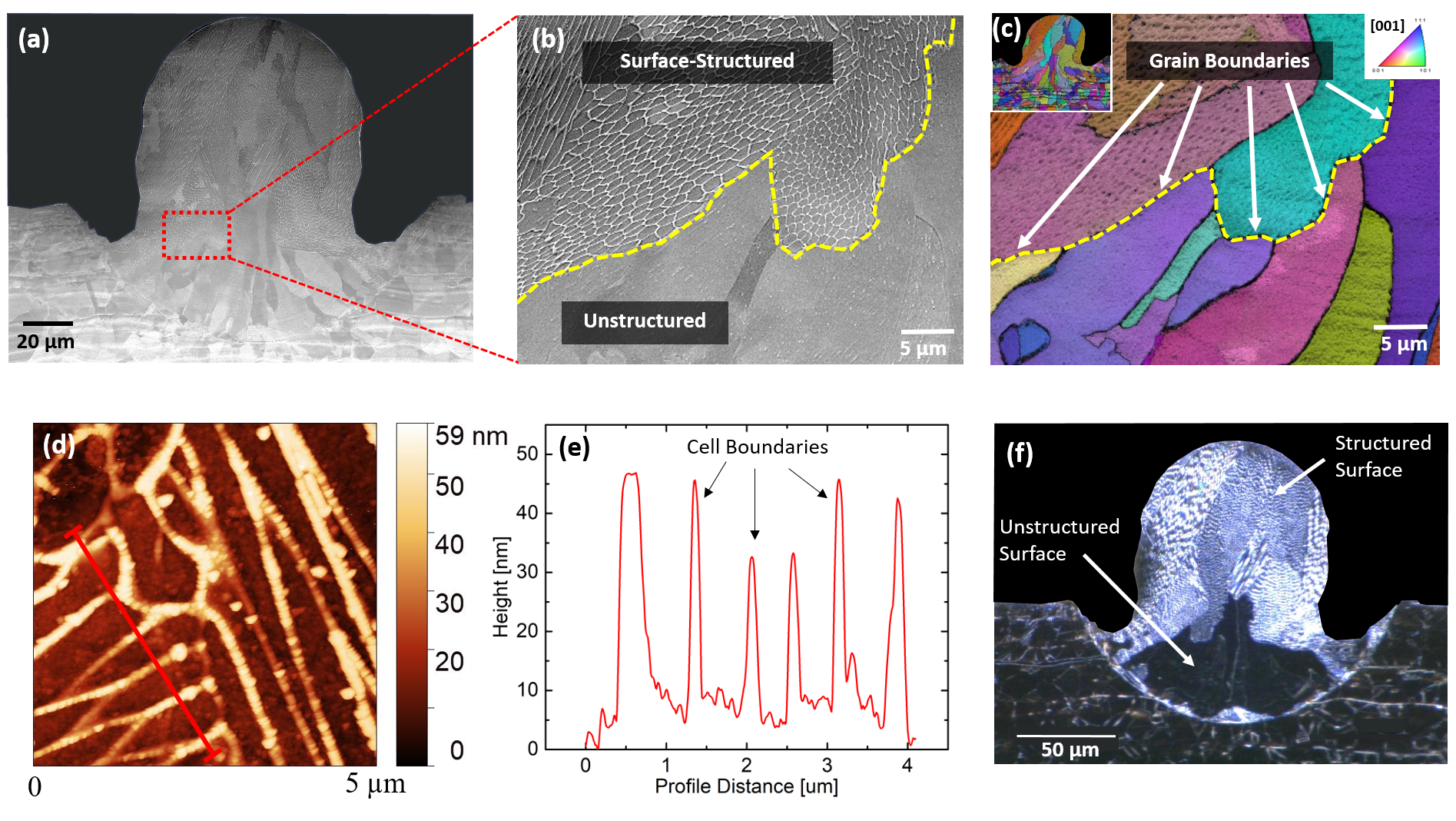}
	\caption{The presence of both regions that do and do not exhibit surface substructure upon etching is observed via (a) low magnification SEM and (b) magnified SEM image clearly delineating the two types of regions (c) is the corresponding area via EBSD-IPF revealing the coincidence of the surface structure/non-surface structure boundary with multiple local grain boundaries, (d) and (e) are AFM topographical maps and corresponding profile demonstrating that the cell walls are protrusions, as opposed to grooves, and (f) is the corresponding dark-field image, clearly depicting the non-surface-structured regions(dark), and those that exhibit surface cell structure upon etching (specular).}\label{fig:surf_v_non}
\end{figure}

Figure \ref{fig:surf_v_non}a presents an SEM image of a polished and etched cross-section of a single track laser-melted powder 316L scan line. It was produced by a Concept Laser M2 SLM system, using factory-default process parameters (for 316L), including laser power, P = 370 W, scan velocity, V = 900 mm/s, and laser spot size d = 160 $\mu m$. The SEM image shown in Fig. \ref{fig:surf_v_non}b clearly reveals the elicitation of surface-cellular microstructures upon etching, while also clearly showing regions that do not exhibit these features. Figure \ref{fig:surf_v_non}c shows an electron backscatter diffraction (EBSD) inverse pole figure (IPF) map from the area corresponding to that shown in Fig. \ref{fig:surf_v_non}b. It clearly reveals that the boundary separating regions that exhibit surface substructure from those that do not, also corresponds to grain boundaries. In fact this was typical of all boundaries separating such regions observed in more than two dozen cross-sectional analyses. This is a critical point, as, if the process (e.g. laser power, spot size and scan velocity) were coincidentally operating near the bounds of a parameter space over which cellular solidification was active, its presence or absence would be due to changes in the local cooling/solidification rate at the advancing solid/liquid interface, and would be independent of grain boundary geometry. Furthermore, prior studies  as well as our own observe the same phenomenon using laser scan rates \cite{Liu2018354, Sun2016197} (an indirect surrogate for cooling rate) up to 7,000 mm/s, well outside of any coincidental parameter-space operation. 

Figures \ref{fig:surf_v_non}d and \ref{fig:surf_v_non}e display a topographical image and corresponding profile obtained via atomic force microscopy (AFM), revealing that cell walls protrude from the surface, implying enhanced chemical stability of cell walls relative to cell interiors. If the cell structures were due to the presence of dislocations alone, then the cell walls would be expected to etch preferentially, and form grooves or pits instead of protrusions \cite{PTmetalloy}. Enhanced chemical stability might be expected upon local solute segregation to cell walls.  Figure \ref{fig:surf_v_non}f presents the corresponding dark-field optical image of the same cross-section, demonstrating the stark difference between surface-structured (diffuse) and non-surface-structured regions (specular/black). This contrast enables an extremely efficient means of image analysis (detailed below). 

Although prior studies, as well as our own observations (see supplement) reveal full infiltration of surface cellular features in bulk AM-produced parts, i.e. many layers, this is not the case for single-tracks. In fact, every single-track cross-section analyzed incorporates some area within the processed zone that yields no visible surface cell structure upon polishing and etching. In some cases almost the entire cross-section appears to be devoid of structure. Though puzzling, the presence of these “non-surface-cell-structured” regions is fortuitous, and greatly facilitates the explanation of the mechanism for cellular formation.

\begin{figure}
	\includegraphics[width=0.99\linewidth]{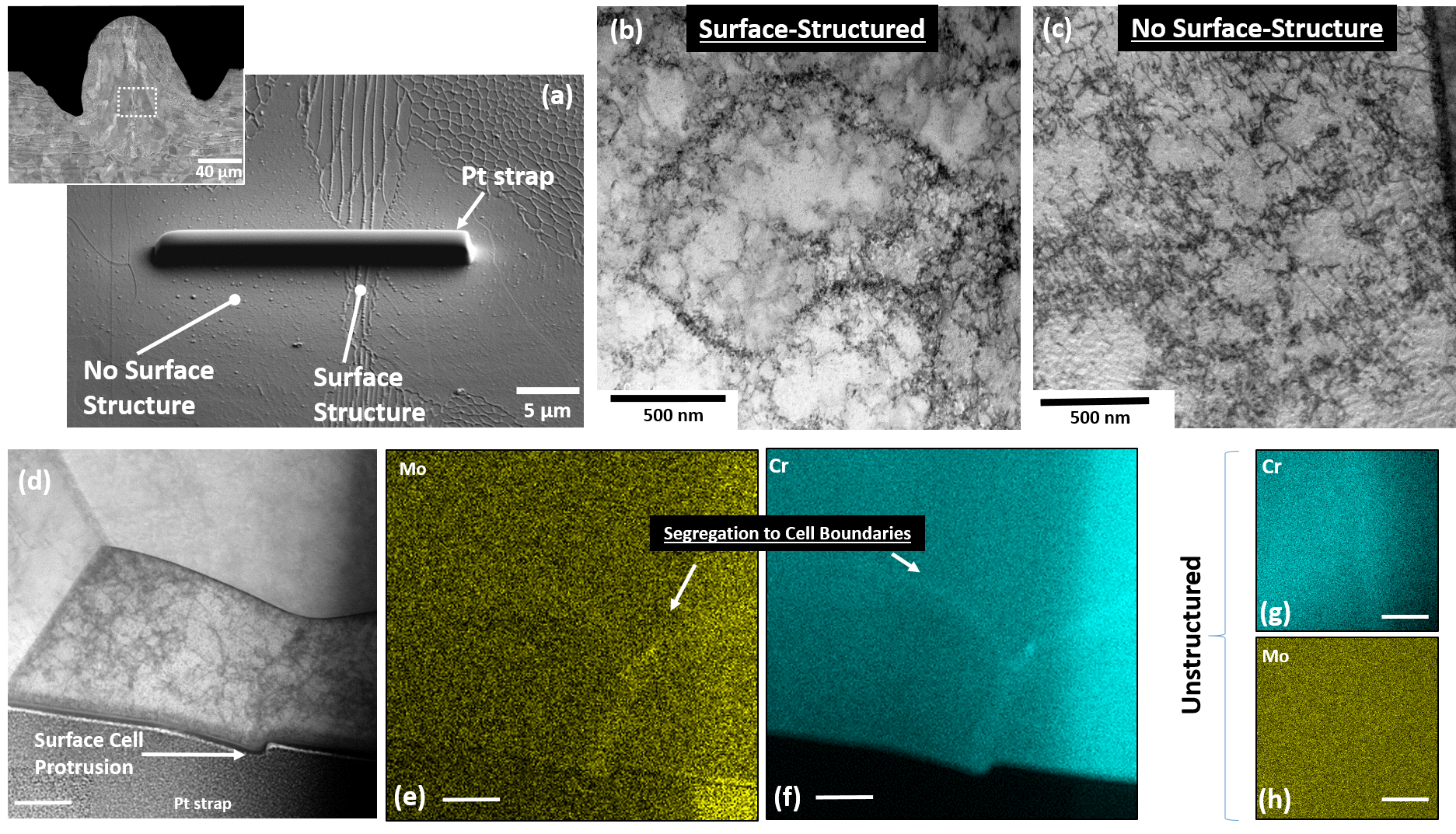}
	
	\caption{Surface and non-surface structured regions were analyzed via TEM. (a) is a STEM image detailing the location of a FIB-prepared sample from a single-track cross-section (inset) spanning a structure/non-structure boundary (b) is a STEM image from a surface-structured region showing dislocation cell structure (c) is from the non-surface structured region also revealing dislocation cell structure, however only the dislocation cell structures in the surface structured region (d-f) (scale bar = 200 nm) exhibit microsegregation to cell walls for a grain that exhibited surface cell structure, while the non-surface structured regions (g,h) (scale bar = 500 nm) remain chemically homogeneous.}\label{fig:TEM}
\end{figure}

Figure \ref{fig:TEM}a shows an SEM image of a location for extracting a TEM foil via focused ion beam (FIB) preparation within a single-track cross-section (inset) specifically chosen for analyzing the differences between crystals exhibiting surface structure, and those that do not. The Pt strap defines the top of the TEM foil. Figures \ref{fig:TEM}b and \ref{fig:TEM}c are the corresponding STEM images from surface and non-surface-structured regions respectively. Figure \ref{fig:TEM}b reveals the well-defined dislocation-tangle cell structures observed previously. Figure \ref{fig:TEM}c also reveals a cellular structure, though more diffuse, and smaller characteristic dimension than its counterpart. Figures \ref{fig:TEM}d-f show a STEM and corresponding EDS maps for Mo and Cr, respectively for a region that clearly exhibited surface cell structure. Segregation to cell walls, including to the surface protrusion that defines the surface cell structure is clearly observable. However, upon a corresponding examination, there is no discernible segregation for the non-surface structured region (Figs. \ref{fig:TEM}g and \ref{fig:TEM}h), indicating that the structures exhibited in Fig. \ref{fig:TEM}c are cells defined solely by dislocations. 

Micro-segregation is a prerequisite for cellular solidification \cite{PTmetalloy}, and is driven by the imbalance in solute solubility at an advancing solid/liquid interface while solidifying. Upon full solidification, solute rejected into the liquid results in compositionally periodic structures. It therefore follows that unless some hitherto unknown solid-state process results in chemical [re]homogenization, the absence of compositional periodicity implies cellular solidification did not occur. It also follows that the formation of the dislocation cell structure (Fig. \ref{fig:TEM}c) is not due to an existing, spatially periodic compositional scaffolding. 

\section*{Cell Formation}
The formation of pure dislocation cell structures in work-hardened metals is a well-known phenomenon  \cite{Holt19703197}, whose dynamics of formation \cite{Kuhlmann-Wilsdorf19837, Galindo-Nava20124370}, associated stress/strain energy fields \cite{Mughrabi19831367}, and geometries  \cite{Kuhlmann-Wilsdorf198279} have been rigorously characterized experimentally, and accompanied by the formulation of underlying theories. In fact, prior work \cite{Michel19731269} analyzing the dislocation microstructure/substructure of uniaxially loaded, elevated temperature (isothermal) 316L reveals the presence of dislocation structures that are  extremely reminiscent in both size and appearance to those observed in SLM-processed material. Furthermore, a prior AM effort performed using electron-beam melting of pure copper  \cite{Ramirez20114088} also reveals near-identical dislocation structures generated upon resolidification. Cellular solidification is extremely unlikely, as there is no solute available in a pure system for redistribution, though they do report the presence of copper oxide precipitates. 

It is well documented that significant residual stresses are generated upon solidification for SLM-processed metals \cite{Mercelis2006254}. It is straightforward to show that for a single track configuration, the unconstrained thermal strain generated upon solidification, is $\epsilon (T)=\alpha(T)\Delta T$ with $\Delta T = T_{m}-T_{r}$, where $T_{m} \sim 1800 \degree K$ is the equilibrium melt temperature of  the alloy, $T_{r} = 303 \degree K$ is room temperature, and $\alpha(T)$ is the temperature-dependent coefficient of thermal expansion. The underlying, unmelted baseplate acts to constrain this shrinkage, thus generating multiaxial, tensile residual stresses throughout the volume of the processed region. It is critical to note that at elevated temperatures, the yield stress of the material is dramatically lower than at room temperature  \cite{Gardner2010634}, thus providing significantly lower resistance to dislocation motion. We posit that it is this combination of thermal strain generation, coupled with the diminished yield stress that results in the formation of the dislocation cell structure that is widely observed. 

If it is accepted that substructure is due solely to dislocation cell generation, the presence of coincident chemical heterogeneity must also be reconciled. Strain-aging, either static or dynamic \cite{VanDenBeukel19821027}, is a well-known metallurgical/thermomechanical phenomenon. It involves either the sequential or simultaneous straining and heating of a material to induce local solute segregation as a means for strengthening the material via solute/dislocation interaction. In the presence of local defects, Fick’s law is insufficient to describe the conditions for diffusion \cite{PTmetalloy}. In fact, it is the tendency to minimize perturbations in chemical potential that is the thermodynamic driving force for equilibrium, not merely gradients in elemental concentration \cite{PTmetalloy}. It is the strain energy field generated around dislocation structures that drives solute diffusion to minimize local free energy via the formation of Cottrell atmospheres \cite{Cottrell194949}, and not vice-versa. The stabilizing effect of the solute acts to pin the dislocations, thus effectively strengthening the material, and, at elevated temperatures and/or strain levels, gives rise to the Portevin-Le Chatelier effect, or the apparent “saw-tooth” behavior in stress-strain response \cite{Kubin1990697}. 

\begin{figure}
	\includegraphics[width=0.99\linewidth]{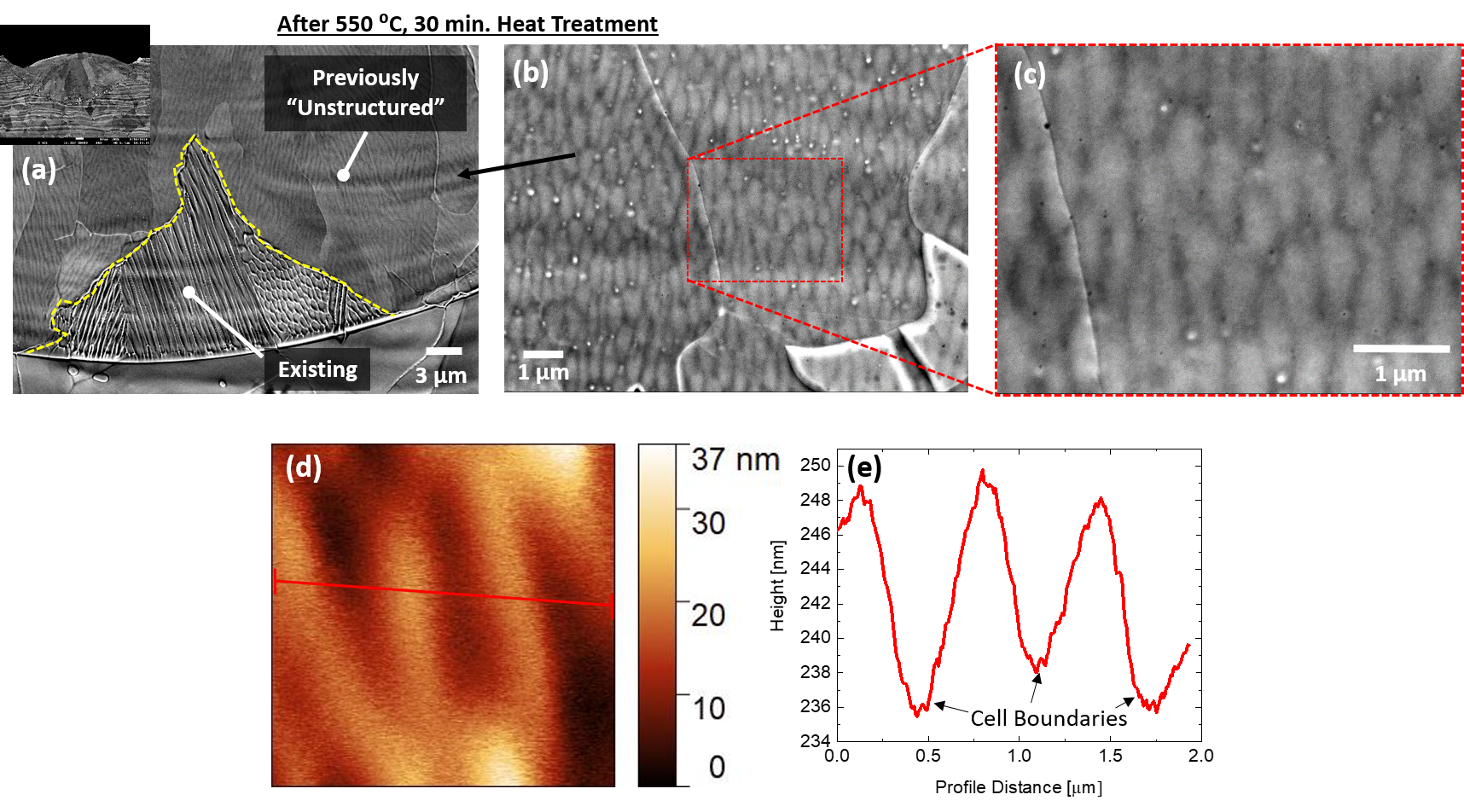}
	\caption{By heat-treating single-track cross-sections (a and inset) at 550 ⁰C for 30 minutes, regions that had previously not exhibited surface cell structure, clearly begin to exhibit cell surface-cell structure via the formation of Cottrell atmospheres, as clearly seen in (b) and (c). (d) and (e) are corresponding AFM scans and profiles, revealing that, as opposed to pre-existing cell structures which produce cell wall protrusions, cell boundaries for transformed yield boundary grooves.}\label{fig:Cottrell}
\end{figure}

We put forth the claim that it is the dynamic evolution of dislocation structures, followed by strain aging that are the underlying mechanisms responsible for the apparent cell structure generation, and subsequent (solid-state) microsegregation. It follows that if the regions that do not exhibit surface cells are purely dislocation structures, then a subsequent heat treatment (well below the melt temperature) may result in local solute diffusion to the cell walls via the formation of Cottrell atmospheres. Indeed, Figs. \ref{fig:Cottrell}a-c reveal just that. Figure 3b shows a magnified region of a single-track cross-section (inset) that was subsequently heat treated at 550 \degree C for 30 minutes. This heat treatment elicits this previously hidden structure (at the surface), though, as seen in the AFM scan and profile in Figs. \ref{fig:Cottrell}d and  \ref{fig:Cottrell}e, these cell walls etch at a greater rate than their interiors, forming grooves instead of protrusions, suggesting that these cell structures are qualitatively different than those observed in Fig. \ref{fig:surf_v_non}. To note, the heat treatment conditions used were informed by prior work \cite{Li2017} demonstrating dynamic strain aging effects in 316L at elevated temperature.

\section*{$\mathbf{\Sigma 3}$ Boundaries and Microsegregation}

\begin{figure}
	\includegraphics[width=0.99\linewidth]{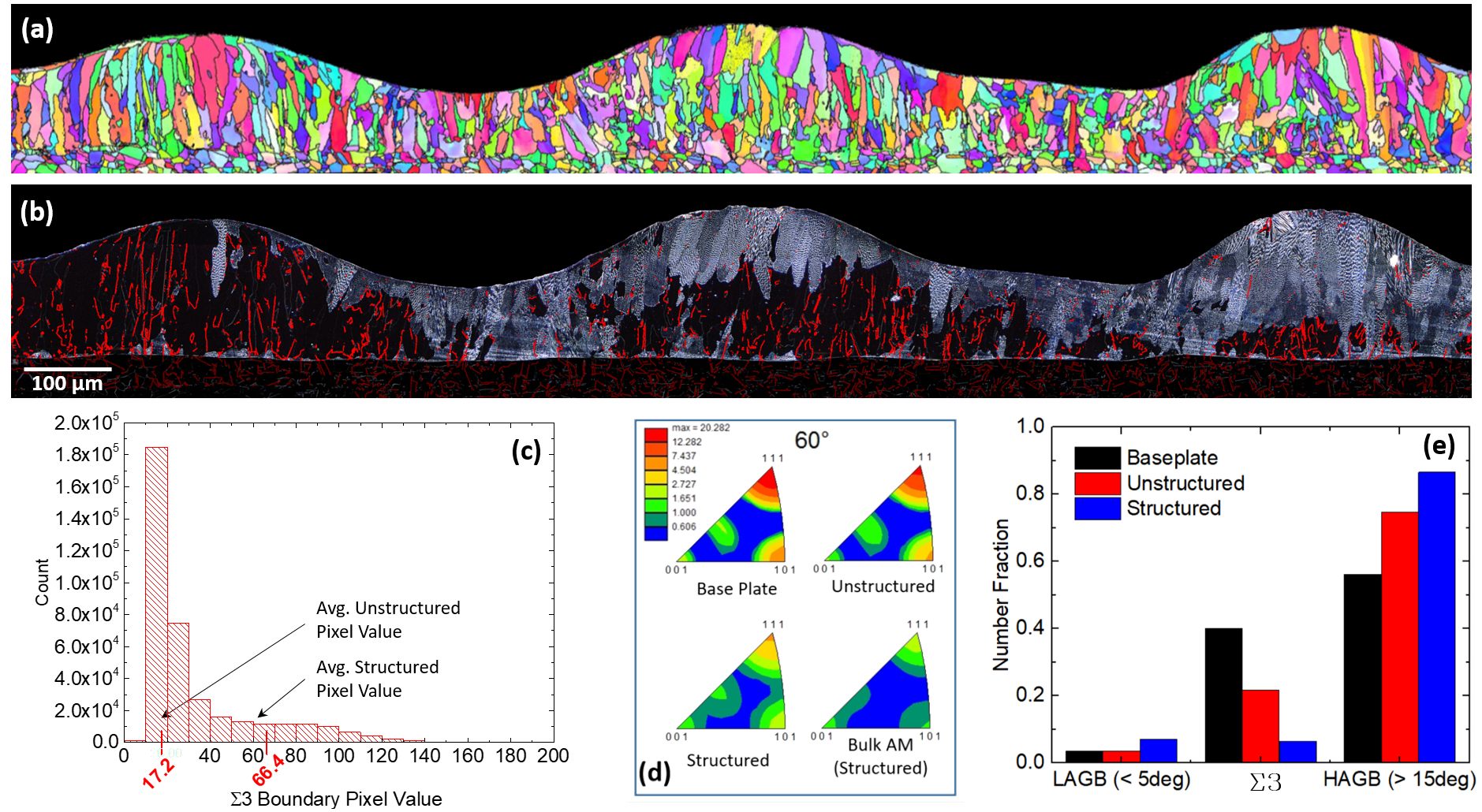}
	\caption{By analyzing a longitudinal section of a single-track, specific correlations between the presence and absence of surface-cell-structure and $\Sigma$3 twin boundaries. (a) is a compilation of EBSD-IPF’s from a ~1 mm length of a single-track deposition (b) is the corresponding dark-field image of the same region, though $\Sigma$3 boundaries determined via (a), superimposed, revealing  a propensity of $\Sigma$3 boundaries in regions not exhibiting surface structure. (c) is a histogram correlating each point on a twin boundary with the values of the coincident pixel neighborhood in the dark field image, (d) is a collection of MDF’s showing the continual reduction in $\Sigma$3 boundaries, and (e) shows the grain boundary character distribution, detailing the evolution in GBCD of SLM material vs. the underlying baseplate.}\label{fig:twins}
\end{figure}

Figure \ref{fig:twins} displays a composite EBSD-obtained IPF map of approximately 1 mm along a scan track sectioned \textit{longitudinally}, i.e. along a face parallel to the laser scan.  It reveals the columnar nature of the growth, as well as clear epitaxial growth from the baseplate at the fusion zone boundary. Figure \ref{fig:twins}b shows the corresponding dark-field optical image revealing regions that exhibit surface structure (diffuse) and lack surface structure (black) upon etching, respectively. Overlaid upon the dark-field image are $\Sigma$3 twin boundaries extracted from the corresponding EBSD analysis (in red). It can be clearly seen that twin boundaries are overwhelmingly present in regions that do \textit{not} exhibit surface cell structure. Figure \ref{fig:twins}c quantifies this qualitative observation by correlating each point on a twin boundary with the values of the coincident pixel neighborhood in the dark field image. This procedure is given in more detail in the methods section. $\Sigma$3 boundary points located within unstructured regions exceed those outside by a full order of magnitude. It is noted that in order to gain sufficient statistics, the data presented in Fig. \ref{fig:twins}c were extracted from approximately 8 mm of scan track, and not just the 1 mm shown in Fig. \ref{fig:twins}b. The full 8 mm track is provided in the supplement. 
Figure \ref{fig:twins}d is a collection of 60\degree  misorientation distribution functions (MDF) for the baseplate, unstructured regions (single track), structured regions (single track), and finally, a bulk AM part. It is seen that there is a significant concentration about the [111] axis, corresponding to $\Sigma$3 boundaries, in the baseplate, which diminishes slightly with respect to the unstructured regions, dramatically with respect to the structured, and almost completely with respect to the bulk AM sample.  Finally, Fig. \ref{fig:twins}e displays so-called grain boundary character distribution (GBCD), which further accentuates the point showing a steep drop-off in the presence of $\Sigma$3 boundaries corresponding with the appearance of surface structure. Notably, no preferred crystallographic orientation was detected for structured or unstructured regions.

The presence of twin boundaries is obviously correlated with the presence or absence of surface structure. However, a causal relationship is less clear. Close inspection of the fusion zone boundary reveals that the twin boundaries that are present in the track are simply those that persist upon epitaxial regrowth into the melt. It can thus be concluded that the (lack of) surface structure is not causing their presence. However, the significant difference in grain boundary energy  \cite{PTmetalloy} between $\Sigma$3 boundaries (19 $mJ/m^2$), and other general high angle grain boundaries (835 $mJ/m^2$) (HAGB) strongly suggests that the overall contribution to the total free energy within a grain may be dramatic, and potentially sufficient for altering the driving force for post-solidification (solid-state) segregation.   

The exact nature of a potential causal relationship is currently incompletely understood, though we offer the following hypothesis. Pure dislocation cell structures are generated upon solidification due solely to local thermomechanical conditions, i.e. orientation with respect to thermal shrinkage and the (local) Schmid factor. The statistical presence or absence of $\Sigma$3 boundaries contribute to the overall free energy of the respective crystals in a manner such that the time at elevated temperature upon cooling from the melt is or is not sufficient for causing solid-state diffusion to occur via the formation of Cottrell atmospheres around the preexisting dislocation cell structures. That is, general high angle grain boundaries contribute significantly to the local volumetric free energy, thus effectively reducing the marginal energy barrier for diffusion. Whereas the presence of $\Sigma$3 boundaries essentially increases the driving force (by lowering the volumetric free energy) for diffusion. Furthermore, HAGB’s are also more efficient sources of dislocation nucleation \cite{Murr20165811}, thus increasing the dislocation density and corresponding strain energy magnitudes associated with cell walls, which in turn can further drive solid-state diffusion.

\section*{Conclusion}
The theory presented accounts for the intermittent reported observation of elemental concentrations as follows: dislocation structures act as sinks for solute diffusion. However the extent of that diffusion will depend on the preponderance of $\Sigma$3 boundaries (and their three-dimensional configuration), elastic strain energy, an overall contribution of average grain boundary energy (due to grain size/total grain boundary area within a representative volume) and spatially localized conditions such as thermal conductivity, which may be decreased or enhanced by combinatorial effects of AM part geometry (and support structures) and the precise laser scan path. Further variables such as the number of heat/reheat cycles (again dependent on scan path) experienced by a particular area will also strongly influence residual strain energy fields. Therefore, depending on the specifics of the AM build and the sampling locations chosen, the extent of observed chemical heterogeneity will vary widely. 

Based on the plurality of evidence, it is clear that cellular solidification is an extremely unlikely candidate for the generation of cellular sub-structuring in AM 316L stainless steel. We have shown that pure dislocation cell structures are indeed present, and it is these structures that drive subsequent solute diffusion to cell walls. Furthermore, the nature of the observed structure is heavily influenced by the presence (or absence) of $\Sigma$3 boundaries.  SLM is essentially providing an intrinsic strain aging treatment that has been shown to significantly enhance mechanical strength \cite{Wang201863}. It is also clear that the extent of this strain aging treatment can, at least in part, account for previously observed mechanical heterogeneities in 316L. Now recognized, this phenomenon can potentially be leveraged to both enhance components, as well as avoid potential deleterious effects stemming from its operation. It can also guide decisions regarding post-process treatments. Furthermore, it would seem somewhat surprising if this phenomenon is limited to austenitic 316L. While previously identified as cellular solidification, it may be potentially fruitful to return to other FCC material systems such as Inconel 718  \cite{Amato20122229}, Al-12Si  \cite{Maity201833}, Al-Si-10Mg \cite{Wu2016311}, Co-29Cr-6Mo \cite{Takaichi201367} and CM247LC Ni-Superalloy \cite{Wang201787} to determine if these similar substructures are in fact due to a form of strain aging. Finally, it is noted that the high solidification rates of SLM imply that strain-rate dependent phenomena may also be playing a vital role, and merit further investigation.

\section*{Methods}
\subsection*{SLM Single Track Generation}
All presented single tracks were generated on a GE/Concept Laser M2 SLM system using P = 370 W, V = 900 mm/s and d = 160 um. Concept Laser CL-20ES 316L stainless steel powder was used with approximate composition Cr17.5 at.\%:Ni11.5 at.\%:Mo2.3 at.\%:Mn1.0at.\%:Si0.5at.\%:Fe Balance (P, C, S $<$ 0.03 at.\%). Powder layers were 25 um thick. All single tracks were processed directly on a 5 mm thick 316L baseplate.

\subsection*{Metallography}
All cross sections and longitudinal sections were cut with a diamond saw, and subsequently ground with successively finer grit 320 -- 1200 SiC paper. This was followed by polishing via 9 $\mu m$, 3 $\mu m$, and 40 $nm$ colloidal silica suspensions. Samples were then etched for five minutes in Kroll’s etchant (DI-H$_20$:H$_2$SO$_4$:HF -- 90:8:2) 

\subsection*{Characterization}

\underline{SEM}: Imaging was performed on a JEOL JSM 7001F scanning electron microscope at 10 kV. 

\noindent
\underline{EBSD}: Crystallographic indexing was performed on a JEOL JSM 7001F scanning electron microscope at 30 kV tilted 70\degree $ $ with respect to the incident electron beam. An EDAX EBSD camera was used with a 0.3 $\mu m$ step size and 4 ms acquisition time.

\noindent
\underline{AFM}: Topographic scans were made with a Veeco atomic force microscope in tapping mode.

\noindent
\underline{TEM/Sample Preparation}: An FEI Scios dual-beam focused ion beam with a Ga ion source operating at 30 kV was used to prepare thin foil samples for analysis. Scanning transmission electron microscopy (STEM) bright field (BF) imaging, and energy dispersive X-ray spectroscopy (EDS) mapping was performed using an aberration-corrected JEOL JEM-ARM200CF operating at 200 kV. 

\subsection*{Heat Treat}
Annealed samples were heat treated in vacuum (10$^{-6}$ Torr) at 550  \degree C for 30 minutes. Samples were heated at a rate of 20 \degree C/min, and subsequently cooled to room temperature over approximately five minutes.

\subsection*{Image Processing}
In order to obtain a common reference for the IPF maps and the corresponding dark field images, we applied a typical image registration procedure \cite{GOSHTASBY1986459,GOSHTASBY1988255}. Image registration is a process where at first, known points are identified on two images and their coordinates are recorded. Subsequently a linear transformation that maps the corresponding points of one image to the other is calculated. Finally one of the images is transformed and both of the images are superimposed on top of each other. In our particular set we identified that an affine transformation is adequate to register the IPF maps and dark fields images to a particularly satisfactory manner. 

The segmentation of the various regions was performed by applying Boolean algebra operations on sets of images.  The sets of four images are shown in Figs. \ref{fig:opt} to \ref{fig:base} (in supplement).

In order to remove very small and equiaxed feature the image that represents the location of the twins is segmented using a connected components algorithm that results in the description of the individual twin components using the procedure outlined in \cite{0201108771}. Subsequently the pixel locations of each of the individual twin image components is used to calculate  the major and minor axes of the ellipse that has the same normalized second central moments as the region ( $l_{I}$ and $l_{II}$ respectively). In addition the area of each of the twin image components $A$ is computed. Subsequently a coverage factor is calculated as:

\begin{equation}
	c = \frac{A}{l_I l_{II}}
\end{equation}
and an equixality factor is computed using the following expression:
\begin{equation}
	q = \frac{l_I}{ l_{II}}
\end{equation}

Finally the set of pixels that belong to components that satisfy the following expression are removed from the twins image ($I){tw}$:
\begin{equation}\label{eq:filter}
	\left\{  2 - q\right\}^0 \land \left\{ c - 0.4 \right\}^0
\end{equation}
and a filtered image $I_{tf}$ is obtained. The curly braces in Eqn. \ref{eq:filter} indicate the Macaulay brackets.

The bead mask is calculated by identifying the location of pixels that are neither at the top nor at the base. In Boolean algebra formalism this can be expressed as:
\begin{equation}
	M_{be} = \lnot M_t \lor \ M_b
\end{equation}

Subsequently, the lightness of the pixels is used to calculate pixels that are white and pixels that are black as follows:
\begin{equation}
	I_w = \left\{  I_o - t \right\}^0 \land M_{be}
\end{equation}

\begin{equation}
	I_b = \left\{  t- I_b \right\}^0 \land M_{be}
\end{equation}

Finally, to calculate the histogram of the pixel values on the darkfield image, we collect the values of pixels that satisfy the condition:
\begin{equation}
	M_{be} \land M_{tf}
\end{equation}

\begin{figure}[h]
	\centering
	\caption{(See supplement) Partial Optical image ($I_o$) after Gaussian filtering used in the identification of structured vs unstructured regions}\label{fig:opt}	
\end{figure}

\begin{figure}[h]
	\centering
	\caption{(See supplement) Partial IPF map in binary format that indicate the location of twin boundaries ($I_{tw}$)}\label{fig:twins2}	
\end{figure}

\begin{figure}[h]
	\centering
	\caption{(See supplement) Binary image indicating the regions above the bead}\label{fig:top}	
\end{figure}

\begin{figure}[h]
	\centering
	\caption{(See supplement) Binary image indicating the regions below the bead}\label{fig:base}	
\end{figure}

%% Put the bibliography here, most people will use BiBTeX in
%% which case the environment below should be replaced with
%% the \bibliography{} command.

%\begin{thebibliography}{1}
%\bibitem{dummy} Articles are restricted to 50 references, Letters
%to 30.
%\bibitem{dummyb} No compound references -- only one source per
%reference.
%\end{thebibliography}
%\section*{References}
%\bibliography{refs}

%% Here is the endmatter stuff: Supplementary Info, etc.
%% Use \item's to separate, default label is "Acknowledgements"

\section*{Acknowledgments}
The authors would like to thank Drs. Edward Gorzkowski, James Wollmershauser and Eric Patterson for their experimental assistance as well as Dr. Noam Bernstein and Dr. David Rowenhorst for their insights. The authors further acknowledge support for this work by the Office of Naval Research through the Naval Research Laboratory's core funding.

%%
%% TABLES
%%
%% If there are any tables, put them here.
%%

\end{document}